\documentclass[letterpaper, 10pt, conference]{ieeeconf}
\IEEEoverridecommandlockouts  

\usepackage{amssymb}
\usepackage{amsmath}
\usepackage{amsfonts}                                
\usepackage[latin1] {inputenc}
\usepackage{enumerate}
\usepackage{mathrsfs}
\usepackage{tabulary}
\usepackage{cite}
\usepackage{psfrag}
\usepackage{cite}
\usepackage{graphicx}          
\usepackage{color}      
\usepackage{epsfig} 
\usepackage{epstopdf}
\usepackage{times} 
\def\qedp{\hspace*{\fill}~{\tiny $\blacksquare$}}
\def\qed{\relax\ifmmode\hskip2em \Box\else\unskip\nobreak\hskip1em $\Box$\fi}

\newtheorem{theorem}{Theorem}
\newtheorem{itlemma}{Lemma}
\newtheorem{itdefinition}{Definition}
\newtheorem{itproposition}{Proposition}
\newtheorem{itresult}{Result}
\newtheorem{itremark}{Remark}
\newtheorem{itassumption}{Assumption}
\newtheorem{itcorollary}{Corollary}
\newtheorem{itexample}{Example}

\newenvironment{remark}{\begin{itremark}\rm}{\end{itremark}}
\newenvironment{assumption}{\begin{itassumption}\rm}{\end{itassumption}}
\newenvironment{lemma}{\begin{itlemma}\rm}{\end{itlemma}}

\title{\LARGE \bf  Towards Stabilization of Distributed Systems under Denial-of-Service} 
 
\author{ Shuai Feng, Pietro Tesi, Claudio De Persis
	\thanks{Shuai Feng, Pietro Tesi and Claudio De Persis are with ENTEG, Faculty of Science and Engineering, University of Groningen, 9747 AG Groningen, The Netherlands
		{\tt\small s.feng@rug.nl, p.tesi@rug.nl, c.de.persis@rug.nl}. }
}

\begin{document}
\maketitle    

\begin{abstract}
In this paper, we consider networked distributed systems in the presence 
of Denial-of-Service (DoS) attacks, namely attacks that prevent transmissions 
over the communication network. First, we consider a simple and typical scenario where communication sequence is purely Round-robin and we explicitly calculate a bound of attack frequency and duration, under which the interconnected large-scale system is asymptotically stable. Second, trading-off system resilience and communication load, we design a hybrid transmission strategy consisting of Zeno-free distributed event-triggered control and Round-robin. We show that with lower communication loads, the hybrid communication strategy enables the systems to have the same resilience as in pure Round-robin. 
\end{abstract}

\section{Introduction}

Cyber-physical systems (CPSs) are increasingly appealing for industry nowadays thanks to the development of computation and communication infrastructures. The application of CPSs ranges from local control systems to large-scale systems, examples being house temperature control systems and regional grid control systems. Owing to the advances in economic and possibly reliability reasons, systems tend to be large-scale, interconnected and spatially distributed, among which communications are operated via wireless network \cite{sandell1978survey}. This triggers the attention towards networked control of large-scale interconnected systems, which are possibly safety-critical and potentially exposed to malicious attacks \cite{7879966}.   

The concept of cyber-physical security mostly concerns security against intelligent attacks. One usually classifies these attacks as either deceptive attacks or Denial-of-Service (DoS). Deceptive attacks affect the trustworthiness of transmitted data \cite{Fawzi, Teixeira2015135}. Instead, DoS compromises the timeliness of information exchange, \emph{e.g.} in the presence of DoS, communications are not possible \cite{xu2005feasibility, sastry}.      

This paper investigates DoS attacks. We consider a large-scale system composed of interconnected subsystems, which are possibly spatially distributed. The information exchange between distributed systems and controllers takes place over a shared communication channel, which implies that all the communication attempts can be denied in the presence of DoS.  

The literature on distributed/decentralized networked control \cite{wang2011event, guinaldo2015distributed, de2011small, de2012event, de2013inter, de2013parsimonious, liu2012decentralized} and centralized system under DoS attacks \cite{ 7574308, sastry,basar, Ugrinovskii, zhang2015optimal,zhang2016optimal , martinez2, CDP:PT:IFAC14,de2015input,dolk,Feng201742, CDP:PT:CDC14, ding2017multi, foroush2012event  } is large and diversified. In \cite{de2013parsimonious}, based on a small-gain approach, the authors propose a parsimonious event-triggered design, which is able to prevent Zeno behavior and stabilize nonlinear distributed systems asymptotically. In \cite{guinaldo2015distributed, de2012event}, event-triggered approaches are discussed within large-scale interconnected systems. By introducing a constant in the triggering condition, the authors prove that the system converges to a region around equilibrium without the occurrence of Zeno behavior.   
In \cite{7574308}, the authors consider a scenario where malicious attacks and genuine packet losses coexist, where the effect of malicious attacks and random packet losses are merged and characterized by an overall packet drop ratio.    
In \cite{ding2017multi}, the authors formulate a two-player zero-sum stochastic game framework to consider a remote secure estimation problem, where the signals are transmitted over a multi-channel network under DoS attacks. 
A problem similar to zero-sum games between controllers and strategic jammers is considered in \cite{Ugrinovskii}.  
In \cite{zhang2016optimal}, the authors investigate DoS from the attacker's viewpoint where the objective is to consume limited energy and maximize the effect induced by DoS attacks. 
The paper \cite{foroush2012event} considers a stabilization problem where transmissions are event-based and the network is corrupted by periodic DoS attacks. 
In \cite{CDP:PT:IFAC14,de2015input}, a framework is introduced where DoS attacks are characterized by \emph{frequency} and \emph{duration}. 
The contribution is an explicit characterization of 
DoS frequency and duration 
under which stability can be preserved
through state-feedback control. 
Extensions have been considered 
dealing with dynamic controllers \cite{dolk, Feng201742} and
nonlinear system \cite{CDP:PT:CDC14}.

In this paper, we consider networked distributed systems under DoS attacks, which has not been investigated so far under the class of DoS attacks introduced in \cite{CDP:PT:IFAC14, de2015input}. Previously in \cite{CDP:PT:IFAC14, de2015input, dolk, Feng201742, CDP:PT:CDC14}, the authors analyze the behavior of systems in a centralized-system manner, where the major characteristic is that all the states are assumed to be collected and sent in one transmission attempt. In this paper, we analyze the problem from the distributed system point of view, where the interconnected subsystems share one communication channel and transmission attempts of the subsystems take place asynchronously. The contribution of this paper is twofold. First, we consider a simple but typical scenario where the communication sequence is purely Round-robin and we explicitly compute a bound on attack frequency and duration, under which the large-scale system is asymptotically stable. Second, trading-off system resilience and communication load, we design a hybrid transmission strategy. Specifically, in the absence of DoS attacks, we design a distributed event-triggered control using small gain argument, which guarantees practical stability of the closed-loop system while preventing the occurrence of Zeno behavior. During DoS-active periods, communication switches to Round-robin, aiming at quick communication restore. This hybrid communication strategy surprisingly ends up with the same bound as pure Round-robin transmission but promotes the possibility to save communication resources. 

The paper is organized as follows. In Section \uppercase\expandafter{\romannumeral 2}, we introduce the framework of interest along with the considered family of DoS attacks. In Section \uppercase\expandafter{\romannumeral 3}, we present the main result. We first use small gain approach to study large-scale system under Round-robin. Subsequently, we introduce the main result of this paper: the characterization of frequency and duration of DoS attacks, under which the large-scale system is asymptotically stable. Section \uppercase\expandafter{\romannumeral 4} briefly introduces a hybrid transmission design, which achieves the same result as in Section \uppercase\expandafter{\romannumeral 3} with lower communication load.  Section \uppercase\expandafter{\romannumeral 5} discusses numerical simulations and Section \uppercase\expandafter{\romannumeral 6} ends the paper with conclusions and possible future research directions.

\subsection{Notation}
We denote by $\mathbb R$ the set of reals. Given 
$\alpha \in \mathbb R$, we let $\mathbb R_{> \alpha}$
($\mathbb R_{\geq \alpha}$) denote
the set of reals greater than (greater than or equal to) $\alpha$.
We let $\mathbb N_0$ denote the set of nonnegative integers, 
$\mathbb N_0 := \{0,1,\ldots\}$. The prime denotes transpose.
Given a vector $v \in \mathbb R^n$, $\|v\|$ is its Euclidean norm. 
Given a matrix $M$, $\|M\|$ is its spectral norm. 
Given two sets $A$ and $B$,
we denote by $B \backslash A$ the relative complement of $A$ in $B$, 
\emph{i.e.}, the set of all elements belonging to $B$, but not to $A$. 

\section{Framework}
\subsection{Networked distributed system }
Consider a large-scale system consisting of $N$ interacting subsystems, whose dynamics satisfy
\begin{eqnarray} {\label{system}}
\dot x_i (t) = A_i x_i(t) + B_i u_i(t) + \sum_{j \in N_i} H_{ij}x_j(t) 
\end{eqnarray}
where $A_i$, $B_i$ and $H_{ij}$ are matrices with appropriate dimensions and $t\in \mathbb{R}_{\ge 0}$. $x_i(t)$ and $u_i(t)$ are state and control input of subsystem $i$, respectively. Here we assume that all the subsystems are full state output. $N_i$ denotes for the set of neighbors of subsystem $i$. Subsystem $i$ physically interacts through $\sum_{j \in N_i}H_{ij}x_j(t)$ with its neighbor subsystem(s) $j \in N_i$. Here we consider bidirectional edges, \emph{i.e.} $j\in N_i$ when $i \in N_j$.

The distributed systems are controlled via a shared networked channel, through which distributed plants broadcast the measurements and controllers send control inputs. The computation of control inputs is based on the transmitted measurements. The received measurements are in sample-and-hold fashion such as $ x_i(t^i _k) $ where $t_k^i$ represents the sequence of transmission instants of subsystem $i$. We assume that there exists a feedback matrix $K_i$ such that $\Phi_i=A_i+B_iK_i$ is Hurwitz. Therefore, the control input applied to subsystem $i$ is given by
\begin{eqnarray} {\label{control input}}
u_i(t) &=& K_i {x}_i(t_k^i) + \sum_{j \in N_i} L_{ij}  x_j(t^j _k) 
\end{eqnarray}
where $L_{ij}$ is the coupling gain in the controller. Here we assume that the channel is noiseless and there is no quantization. Moreover, we assume that the network transmission delay and the computation time of control inputs are zero.

\subsection{DoS attacks--frequency and duration}

We refer to Denial-of-Service as the phenomenon for which transmission attempts may fail. In this paper, we do not distinguish between transmission failures due to channel unavailability and transmission failures because of DoS-induced packet corruption. Since the network is shared, DoS simultaneously affects the communication attempts of all the subsystems.

Clearly, the problem in question does not have a solution 
if the DoS amount is allowed to be arbitrary.
Following \cite{de2015input}, we consider a general DoS model
that constrains the attacker action in time 
by only posing limitations on the frequency of DoS attacks and their duration.
Let 
$\{h_n\}_{n \in \mathbb N_0}$, $h_0 \geq 0$, denote the sequence 
of DoS \emph{off/on} transitions, \emph{i.e.},
the time instants at which DoS exhibits 
a transition from zero (transmissions are possible) to one 
(transmissions are not possible).
Hence,
\begin{eqnarray}  \label{DoS_intervals}
H_n :=\{h_n\} \cup [h_n,h_n+\tau_n[  
\end{eqnarray}
represents the $n$-th DoS time-interval, of a length 
$\tau_n \in \mathbb R_{\geq 0}$,
over which the network is in DoS status. If $\tau_n=0$, then
$H_n$ takes the form of a single pulse at $h_n$.  
If $\tau_n \ne 0$, $[h_n,h_n+\tau_n[$ represents an interval from the instant $h_n$ (include $h_n$) to ($h_n+\tau_n)^-$ (arbitrarily close to but exclude $h_n+\tau_n$). Similarly, $[\tau, t[$ represents an interval from $\tau$ to $t^-$. 
Given $\tau,t \in \mathbb R_{\geq0}$ with $t\geq\tau$, 
let $n(\tau,t)$
denote the number of DoS \emph{off/on} transitions
over $[\tau,t[$, and let 
\begin{eqnarray}  \label{DoS_intervals_union}
\Xi(\tau,t) := \bigcup_{n \in \mathbb N_0} H_n  \, \bigcap  \, [\tau,t] 
\end{eqnarray}
denote the subset of $[\tau,t]$ where the network is in DoS status. The subset of time where DoS is absent is denoted by
\begin{eqnarray}
\Theta(\tau,t) :=[\tau,t] \setminus \Xi(\tau,t) 
\end{eqnarray}

We make the following assumptions. 

\begin{assumption}
	(\emph{DoS frequency}). 
	There exist constants 
	$\eta \in \mathbb R_{\geq 0}$ and 
	$\tau_D \in \mathbb R_{> 0}$ such that
	\begin{eqnarray} \label{ass:DoS_slow_frequency} 
	n(\tau,t)  \, \leq \,  \eta + \frac{t-\tau}{\tau_D}
	\end{eqnarray}
	for all  $\tau,t \in \mathbb R_{\geq0}$ with $t\geq\tau$.
	\qedp
\end{assumption}

\begin{assumption} 
	(\emph{DoS duration}). 
	There exist constants $\kappa \in \mathbb R_{\geq 0}$ and $T  \in \mathbb R_{>1}$ such that
	\begin{eqnarray} \label{ass:DoS_slow_duration}
	|\Xi(\tau,t)|  \, \leq \,  \kappa + \frac{t-\tau}{T}
	\end{eqnarray}
	for all  $\tau,t \in \mathbb R_{\geq0}$ with $t\geq\tau$. 
	\qedp
\end{assumption}

\begin{remark}
	Assumptions 1 and 2
	do only constrain a given DoS signal in terms of its \emph{average} frequency and duration.
	Actually, $\tau_D$ can be defined as the average dwell-time between 
	consecutive DoS off/on transitions, while $\eta$ is the chattering bound.
	Assumption 2 expresses a similar 
	requirement with respect to the duration of DoS. 
	It expresses the property that, on the average,
	the total duration over which communication is 
	interrupted does not exceed a certain \emph{fraction} of time,
	as specified by $1/T$.
	Like $\eta$, the constant $\kappa$ plays the role
	of a regularization term. It is needed because
	during a DoS interval, one has $|\Xi(h_n,h_n+\tau_n)| = \tau_n >  \tau_n /T$.
	Thus $\kappa$ serves to make (\ref{ass:DoS_slow_duration}) consistent. 
	Conditions $\tau_D>0$ and $T>1$ imply that DoS cannot occur at an infinitely
	fast rate or be always active. \qedp
\end{remark}

\section{Main result}

In this section, our objective is to find stability conditions for the networked distributed systems under DoS attacks. We first study the stabilization problem of large-scale systems under a digital communication channel in the absence of DoS.

\subsection{A small-gain approach for large-scale systems under networked communication}

For each subsystem $i$, we denote by $e_i(t)$ the error between the value of the state transmitted to its neighbors and the current state, \emph{i.e.},
\begin{eqnarray} \label{definition of error}
e_i(t) =  x_i(t_k^i)- x_i(t), \quad i=1, 2, ..., N 
\end{eqnarray}
Then combine (\ref{system}), (\ref{control input}) and (\ref{definition of error}), the dynamics of subsystem $i$ can be written as
\begin{eqnarray} \label{complete system dynamics}
\dot x_i (t) &=& \Phi_i x_i(t) + B_iK_i e_i(t)+ \sum_{j \in N_i} (B_i L_{ij} + H_{ij}) x_j(t)  \nonumber \\
&&+   B_i \sum_{j \in N_i} L_{ij} e_j(t)
\end{eqnarray}
from which one sees that the dynamics of subsystem $i$ depend on the interconnected neighbors $x_j(t)$ as well as $e_i(t)$, $e_j(t)$ and the coupling parameters. Intuitively, if the couplings are weak and $e$ remains small, then stability can be achieved. Here, the notion ``smallness" of $e$ can be characterized by the $x$-dependent bound $\|e_i(t)\| \le \sigma_i  \|x_i(t)\|$,
in which $\sigma_i$ is a suitable design parameter. Notice that this is not the network update rule. 

We implement a periodic sampling protocol, \emph{e.g.} Round-robin, as our update law. In this respect, we make the following hypothesis.

\begin{assumption}
	(\emph{Inter-sampling of Round-robin}). 
	In the absence of DoS attacks, there exists an inter-sampling interval $\Delta$ such that 
\begin{eqnarray} {\label{error bound}}
\|e_i(t)\| \le \sigma_i  \|x_i(t)\|
\end{eqnarray}
holds, where $\sigma_i$ is a suitable design parameter.	\qedp
\end{assumption}

For centralized settings, values of $\Delta$ satisfying a bound like ({\ref{error bound}}) can be explicitly determined. On the other hand, in \cite{de2013inter, mazo2011decentralized}, the authors compute and apply a lower bound of time elapsed between two events to prevent Zeno behavior, where the distributed/decentralized systems are asymptotically stable. The problem of obtaining $\Delta$ is left for future research.

As mentioned in the foregoing argument, $\sigma_i$ should be designed carefully. Otherwise, even if there exists a $\Delta$ under which (\ref{error bound}) holds, in the event of an inappropriate $\sigma_i$, stability can be lost as well. 

Given any symmetric positive definite matrix $Q_i$, let $P_i$ be the unique solution of the Lyapunov equation $\Phi_i^TP_i+P_i\Phi_i+Q_i=0$. For each $i$, consider the Lyapunov function $V_i=x_i^TP_ix_i$, which satisfies
\begin{eqnarray}
\lambda_{\min}(P_i) \|x_i(t)\|^2 \le V_i(x_i(t)) \le \lambda_{\max}(P_i) \|x_i(t)\|^2
\end{eqnarray}
where $\lambda_{\min}(P_i)$ and $\lambda_{\max}(P_i)$ represent the smallest and largest eigenvalue of $P_i$, respectively.
The following lemma presents the design of $\sigma_i$ guaranteeing stability.

\begin{lemma}
Consider a distributed system as in (\ref{system}) along with a control input as in (\ref{control input}). Suppose that the spectral radius $r(A^{-1}B)<1$. The distributed system is asymptotically stable if $\sigma_i$ satisfies
\begin{eqnarray}
\sigma_i &<& \sqrt{\frac{l_i}{j_i}}
\end{eqnarray}
where $l_i$ is the $i$-th entry of row vector $L:=\mu^T(A-B)=[l_1, \,l_2, ..., l_N]$ and $j_i$ is the $j$-th entry of row vector $J:=\mu^T \Gamma=[j_1, \,j_2, ..., j_N]$. $\mu\in \mathbb{R}^N_+$ is an arbitrary column vector satisfying $\mu^T(-A+B)<0$. The matrices $A$, $B$ and $\Gamma$ are
given by  
\begin{eqnarray}
&&A=\left[ \begin{array}{ccc}  \alpha_1  &  \,  & \, \\ \, & \ddots & \,\\ \, & \, & \alpha_N \end{array} \right] \\
&&B=\left[ \begin{array}{cccc}  0  & \beta_{12}  & \cdots & \beta_{1N} \\ \beta_{21} & 0 &\beta_{23} & \beta_{2N} \\ \vdots  & \vdots & 0 & \vdots\\ \beta_{N1}  &\beta_{N2} & \cdots & 0 \end{array} \right] \\
&&\Gamma=\left[ \begin{array}{cccc}  \gamma_{11}  & \gamma_{12}  & \cdots & \gamma_{1N} \\ \gamma_{21} & \gamma_{22} &\gamma_{23} & \gamma_{2N} \\ \vdots  & \vdots & \vdots & \vdots\\ \gamma_{N1}  &\gamma_{N2} & \cdots & \gamma_{NN} \end{array} \right]
\end{eqnarray}
with 
\begin{eqnarray}
\alpha_i &=& \lambda_{\min} (Q_i) -  \delta - \sum_{j \in N_i} 2\delta \\
\beta_{ij}    &=&  \frac{\|P_i\|^2\|B_iL_{ij}+H_{ij}\|^2}{\delta}\\
\gamma_{ii} &=&  \frac{\|P_i\|^2\|B_iK_i\|^2}{\delta} \\
\gamma_{ij}    &=&  \frac{\|P_i \|^2\|B_iL_{ij}\|^2}{\delta}
\end{eqnarray}
where $\delta$ is a positive real such that $\alpha_i >0$ and $\lambda_{\min}(Q_i)$ is the smallest eigenvalue of $Q_i$  for $i=1,2, ..., N$. 
\end{lemma}

\emph{Proof}. Recalling that $V_i=x_i^TP_ix_i$, the derivative of $V_i$ along the solution to (\ref{complete system dynamics}) satisfies
\begin{eqnarray}
\dot V_i (x_i(t)) &\le&  -\lambda_{\min}(Q_i) \|x_i(t)\|^2   \nonumber\\
			&&+ \|2P_iB_iK_i\| \|x_i(t)\| \|e_i(t)\| \nonumber\\
			&&+ \sum_{j\in N_i} \|2P_i(B_iL_{ij}+H_{ij})\|\|x_i(t)\|\|x_j(t)\| \nonumber \\
		    &&+  \sum_{j\in N_i} \|2P_iB_iL_{ij}\|\|x_i(t)\|\|e_j(t)\|
\end{eqnarray}
Observe that for any positive real $\delta$, the Young's inequalities yield
\begin{eqnarray}
&&\|2P_iB_iK_i\| \|x_i(t)\| \|e_i(t)\| \nonumber\\
&\le& \delta \|x_i(t)\|^2 + \frac{\|P_i\|^2  \|B_iK_i\|^2}{\delta} \|e_i(t)\|^2  \\
&&\|2P_i(B_iL_{ij}+H_{ij})\|\|x_i(t)\|\|x_j(t)\| \nonumber \\
&\le& \delta \|x_i(t)\|^2 + \frac{\|P_i\|^2\|B_iL_{ij}+H_{ij}\|^2}{\delta} \|x_j(t)\|^2  \\
&&\|2P_iB_iL_{ij}\|\|x_i(t)\|\|e_j(t)\| \nonumber\\
&\le& \delta \|x_i(t)\|^2 +  \frac{\|P_i\|^2\|B_iL_{ij}\|^2}{\delta} \|e_j(t)\|^2 
\end{eqnarray}
Hence, the derivative of $V_i$ along the solution to (\ref{complete system dynamics}) satisfies
\begin{eqnarray}\label{raw derivative Lyapunov}
\dot V_i (x(t)) &\le& -\alpha_i\|x_i(t)\|^2  + \sum_{j\in N_i} \beta_{ij} \|x_j(t)\|^2  \nonumber\\
		&& + \gamma_{ii} \|e_i(t)\|^2 + \sum_{j\in N_i} \gamma_{ij} \|e_j(t)\|^2 
\end{eqnarray}
where $\alpha_i$, $\beta_{ij}$, $\gamma_{ii}$ and $\gamma_{ij}$ are as in Lemma 1. Notice that one can always find a $\delta$ such that $\alpha_i > 0$ for $i=1, 2, ..., N$. 

By defining vectors
\begin{eqnarray}
V_{vec}(x_i(t))&:=&[V_1(x_1(t)), V_2(x_2(t)),..., V_N(x_N(t))]^T \nonumber\\
\|x(t)\|_{vec} &:=& [\|x_1(t)\|^2, \|x_2(t)\|^2, ..., \|x_N(t)\|^2]^T \nonumber\\
\|e(t)\|_{vec} &:=& [\|e_1(t)\|^2, \|e_2(t)\|^2, ..., \|e_N(t)\|^2]^T \nonumber
\end{eqnarray}
the inequality (\ref{raw derivative Lyapunov}) can be compactly written as 
\begin{eqnarray}
\dot V_{vec} (x_i(t)) \le (-A+B) \|x(t)\|_{vec} + \Gamma \|e(t)\|_{vec}
\end{eqnarray}
with $A$, $B$ and $\Gamma$ being as in Lemma 1. 

If the spectral radius satisfies $r(A^{-1}B)<1$, there exists a positive vector $\mu \in \mathbb{R}_+ ^ n$ such that $\mu^T (-A + B)<0$. We refer readers to \cite{dashkovskiy2011small} for more details. We select the Lyapunov function $V(x(t)):=\mu^T V_{vec}(x_i(t))$. Then the derivative of $V$ yields
\begin{eqnarray}
\dot V (x(t))&=& \mu^T \dot V_{vec}(x_i(t))  \nonumber \\
		&\le& \mu^T (-A+B) \|x(t)\|_{vec} + \mu^T \Gamma \|e(t)\|_{vec} \nonumber \\
\end{eqnarray} 
By noticing that $\mu^T (-A+B)<0$, we have 
\begin{eqnarray} \label{dot V in vector form}
\dot V (x(t)) \le -L  \|x(t)\|_{vec} + J \|e(t)\|_{vec}
\end{eqnarray} 
where $L: = \mu^T (A-B)$ and $J:=\mu^T \Gamma$ are row vectors. Denote $l_i$ and $j_i$ as the entries of $L$ and $J$, respectively. Then, (\ref{dot V in vector form}) yields 
\begin{eqnarray} \label{dot V}
\dot V(x(t)) &\le& -\sum_{i\in N} l_i \|x_i(t)\|^2 + \sum_{i \in N}j_i \|e_i(t)\|^2 \nonumber\\
		&=&  -\sum_{i \in N}(l_i \|x_i(t)\|^2 - j_i \|e_i(t)\|^2)
\end{eqnarray}
which implies asymptotic stability with $\sigma_i < \sqrt{\frac{l_i}{j_i}}$.
\qedp 
\begin{remark}
	Lemma 1 can only deal with the case where $j_i >0$. The case $j_i=0$ is only possible whenever every entry in the column $i$ of $\Gamma$ is zero. In fact, $j_i=0$ implies that the error $\|e_i(t)\|$ never contributes to the system dynamics via (\ref{dot V}), which in turn implies that $\|e_i(t)\|$ does not affect stability at all. Therefore, in the case $j_i=0$, no constraint on $\|e_i(t)\|$ is imposed. \qedp  
\end{remark}

\subsection{Stabilization of distributed systems under DoS}
In the previous analysis, we have introduced the design of a suitable $\sigma_i$ and hence error bound, under which the system is asymptotically stable in the absence of DoS. By hypothesis, we also assumed the existence of a Round-robin transmission that satisfies such error bound. In the presence of DoS, (\ref{error bound}) is possibly violated even though the sampling strategy is still Round-robin. Under such circumstances, stability can be lost. Hence, we are interested in the stabilization problem when the Round-robin network is under DoS attacks.

\begin{theorem}
Consider a distributed system as in (\ref{system}) along with a control input as in (\ref{control input}). The plant-controller information exchange takes place over a shared network, in which the communication protocol is Round-robin with sampling interval $\Delta$ as in Assumption 3. The large-scale system is asymptotically stable  for any DoS sequence satisfying Assumption 1 and 2 with arbitrary $\eta$ and $\kappa$, and with $\tau_D$ and $T$ if
\begin{eqnarray} \label{Theorem 1}
\frac{1}{T}+\frac{\Delta_*}{\tau_D}<\frac{\omega_1}{\omega_1+\omega_2}
\end{eqnarray}
in which $\Delta_*=N\Delta$, $\omega_1:= \min\{\frac{l_i - \sigma_i^2 j_i}{\lambda _{\max} (P_i) \mu_i}\} $ and  $\omega_2:=\frac{4\max\{j_i\}}{\min\{\mu_i\lambda_{\min}(P_i)\}}$. $l_i$, $j_i$, $\mu_i$ and $\sigma_i$ are as in Lemma 1.
\end{theorem}

\emph{Proof.} The proof is divided into three steps:

\emph{Step 1. Lyapunov function in DoS-free periods}. 
In DoS-free periods, by hypothesis of Assumption 3, (\ref{error bound}) holds true with $\sigma_i$ as in Lemma 1 and (\ref{dot V}) is negative. Therefore, the derivative of the Lyapunov function satisfies 
\begin{eqnarray}
\dot V(x(t)) &\le& -\sum_{i \in N}	 (l_i-j_i \sigma_i^2)\|x_i(t)\|^2  \nonumber \\
	   &\le& -\sum_{i \in N}  \frac{l_i - \sigma_i^2 j_i}{\lambda _{\max} (P_i) \mu_i} \mu_i V_i  \nonumber \\
	   &=&		-\omega_1  V  
\end{eqnarray}  
where $\omega_1:= \min\{\frac{l_i - \sigma_i^2 j_i}{\lambda _{\max} (P_i) \mu_i}\} $.
Thus for $t \in [h_n+\tau_n, h_{n+1}[$ (DoS-free time), the Lyapunov function yields
\begin{eqnarray} {\label{Lyaponov out of DoS}}
V(x(t)) \le e^{-\omega_1(t-h_n-\tau_n)}V(x(h_n+\tau_n)) 
\end{eqnarray}

\emph{Step 2. Lyapunov function in DoS-active periods}. Here we let $z_m^i$ denote the last successful sampling instant before the occurrence of DoS. 
Recalling the definition of $e_i(t)$, we obtain that
\begin{eqnarray}
e_i(t) = x_i(z_m^i) - x_i(t) = x_i(h_n) - x_i(t)
\end{eqnarray}
and  
\begin{eqnarray}
\|e_i(t)\|^2 \le \|x_i(h_n)\|^2 + 2 \|x_i(t)\| \|x_i(h_n)\| + \|x_i(t)\|^2
\end{eqnarray}
for $t\in H_n$. By summing up $\|e_i(t)\|^2$ for $i \in N$, we obtain
\begin{eqnarray}
\sum_{i\in N} \|e_i(t)\|^2  &\le & \sum_{i\in N}\|x_i(h_n)\|^2  
							 + \sum_{i\in N} \|x_i(t)\|^2   \nonumber \\
							&&+ \sum_{i\in N} (\|x_i(h_n)\| ^2+ \|x_i(t)\|^2) \nonumber \\
							&=& 2 \sum_{i\in N}\|x_i(h_n)\|^2 + 2\sum_{i\in N}  \|x_i(t)\|^2  
\end{eqnarray} 
If $\sum_{i\in N}\|x_i(h_n)\|^2 \le \sum_{i\in N}\|x_i(t)\|^2 $, we have that $\sum_{i\in N} \|e_i(t)\|^2 \le 4\sum_{i\in N}\|x_i(t)\|^2$. Otherwise, we have $\sum_{i\in N} \|e_i(t)\|^2 \le 4  \sum_{i\in N}\|x_i(h_n)\|^2$

Recalling (\ref{dot V}), it is simple to see that 
\begin{eqnarray}
\dot V (x(t)) &\le& \sum_{i \in N} j_i \|e_i(t)\|^2 
\end{eqnarray}
Thus, for all $t\in H_n$ (DoS-active time) in the case that $\sum_{i\in N}\|x_i(h_n)\|^2 \le \sum_{i\in N}\|x_i(t)\|^2 $, the derivative of the Lyapunov function yields
\begin{eqnarray} \label{Lyaponov in DoS 1}
\dot V(x(t))  &\le& \max\{j_i\} \sum_{i \in N}   \|e_i(t)\|^2 \nonumber\\
&\le&  4 \max\{j_i\}  \sum_{i\in N}\|x_i(t)\|^2  \nonumber\\
&\le& \frac{4\max\{j_i\}}{\min\{\mu_i\lambda_{\min}(P_i)\}}\sum_{i\in N} \mu_iV(x_i(t))\nonumber\\
&=& \omega_2 V(x(t))
\end{eqnarray}
with $\omega_2:=\frac{4\max\{j_i\}}{\min\{\mu_i\lambda_{\min}(P_i)\}}$ . On the other hand, for all  $t\in H_n$ such that $\sum_{i\in N}\|x_i(h_n)\|^2 > \sum_{i\in N}\|x_i(t)\|^2 $, one has
\begin{eqnarray} \label{Lyaponov in DoS 2}
\dot V(x(t))  \le \omega_2 V(x(h_n))
\end{eqnarray}
Thus, (\ref{Lyaponov in DoS 1}) and (\ref{Lyaponov in DoS 2}) imply the Lyapunov function during $H_n$ satisfies
\begin{eqnarray} \label{Lyaponov in DoS 3}
V(x(t))  \le e^{\omega_2(t- h_n)}V(x(h_n))
\end{eqnarray}

\emph{Step 3. Switching between stable and unstable modes.}
Consider a DoS attack with period $\tau_n$, at the end of which the overall system has to wait an additional period with length $N\Delta$ to have a full round of communications. Hence, the period where at least one subsystem transmission is not successful can be upper bounded by $\tau_n+ N \Delta$. For all $\tau, t\in \mathbb{R}_{\ge 0}$ with $t \ge \tau$, the total length where communication is not possible over $[\tau, t[$, say $|\bar{\Xi} (\tau,t)|$, can be upper bounded by 
\begin{eqnarray}
|\bar  \Xi (\tau,t)| &\le& |\Xi (\tau,t)| + (1+ n(\tau,t)) \Delta _* \nonumber \\
					 &\le& \kappa_* + \frac{t-\tau}{T_*}
\end{eqnarray} 
where $\Delta_*=N\Delta$, $\kappa:= \kappa+(1+\eta)\Delta_*$ and $T_*:= \frac{\tau_DT}{\tau_D+T\Delta_*}$. 
Considering the additional waiting time due to Round-robin, the Lyapunov function in (\ref{Lyaponov out of DoS}) yields $V(x(t)) \le e^{-\omega_1(t-h_n-\tau_n-N\Delta)}V(h_n+\tau_n+N\Delta))$ for $t\in[ h_n+\tau_n+N\Delta, h_{n+1}[$ and $V(x(t)) \le e^{\omega_2(t-h_n)}V(h_n)$ for $t\in[ h_n, h_n+\tau_n+N\Delta[$.

Thus, the overall behavior of the closed-loop system can be regarded as a switching system with two modes. Applying simple iterations to the Lyaponov functions in and out of DoS status, one has
\begin{eqnarray} \label{Lyapunov function of switching system 1}
V(x(t)) &\le& e^{-\omega_1   |\bar \Theta (0,t)|} e^{\omega_2   |\bar \Xi (0,t)|} V (x(0)) \nonumber\\
		&\le& e^{\kappa_*(\omega_1+\omega_2)} e^{-\beta_* t} V(x(0)) 
\end{eqnarray}
where $\beta_* :=  \omega_1 -(\omega_1+\omega_2)(\frac{\Delta_*}{\tau_D}+\frac{1}{T}) $. 
By constraining $\beta_* < 0$, one obtains the desired result in (\ref{Theorem 1}). Hence, stability is implied at once. 
\qedp

\begin{remark} \label{remark about final result}
The resilience of the distributed systems depends on the largeness of $\omega_1$ and the smallness of $\omega_2$. To achieve this, one can try to find $K_i$ and $L_{ij}$ such that $\|B_iK_i\|$ and $\|B_iL_{ij}\|$ are small. On the other hand, the sampling interval of Round-robin also affects stability in the sense that it determines how fast the overall system can restore the communication. One can always apply smaller Round-robin inter-sampling time to reduce the left-hand side of (\ref{Theorem 1}) at the expense of higher communication load. 
\end{remark}


\section{Approximation of resilience with reduced communication: hybrid transmission strategy }

In the foregoing argument (\emph{cf. Remark \ref{remark about final result}}), we have shown that system resilience depends on the sampling rate of Round-robin. The faster the sampling rate of Round-robin, the quicker the overall system restores the communication. On the other hand, in DoS-free periods, we are interested in the possibility of reducing communication load while maintaining the comparable robustness as in Section III. To realize this, we propose a hybrid transmission strategy: in the absence of DoS, the communications of the distributed systems are event-based; if DoS occurs, the communications switch to Round-robin until the moment where every subsystem has one successful update.

The advantage of event-triggered control is saving communication resources. However, the effectiveness of prolonging transmission intervals, in turn, appears to be a disadvantage in the presence of DoS. The main shortcoming concerns that event-triggered control could potentially prolong DoS status. For example, consider that the sampling strategy is purely event-based. After a DoS attack, there is a short period where communications are possible, during which the error bounds as in (\ref{error bound}) are not violated so that systems do not update. If DoS appears soon, this is equivalent to the scenario that systems face a longer DoS attack. This indicates that a better strategy is to save communications in the absence of DoS and restore communications as soon as possible when DoS is over, which leads indeed to a hybrid communication strategy. 

\subsection{Zeno-free event-triggered control of distributed systems in the absence of DoS}
Abusing the notation, in this section we denote $\{t_k^i\}$ as the triggering time sequence of subsystem $i$ under event-triggered control scheme. For a given initial condition $x_i(0)$, if $t^i _k$ converges to a finite $t^{i*}$, we say that the event-triggered control induces Zeno behavior \cite{de2013inter, de2013parsimonious}. Hence, Zeno-freeness implies an event-triggered control scheme preventing the occurrence of Zeno behavior. The following lemma addresses the Zeno-free event-triggered control. 
\begin{lemma}
Consider a distributed system as in (\ref{system}) along with a control input as in (\ref{control input}). Suppose that the spectral radius $r(A^{-1}B)<1$. In the absence of DoS, the distributed system is practically stable and Zeno-free if the event-triggered law satisfies
	\begin{eqnarray}{\label{error bound Zeno free}}
	\|e_i(t)\| \le \max\{\sigma_i \|x_i(t)\|, \, c_i\}
	\end{eqnarray}
	in which $c_i$ is a positive finite real and  
	\begin{eqnarray}
	\sigma_i &<& \min\{\sqrt{\frac{l_i}{j_i}} , 1\}
	\end{eqnarray}
	where $l_i$ and $j_i$ are the same as in Lemma 1.
\end{lemma}

\emph{Proof}. 
By Lemma 1 if spectral radius $r(A^{-1}B)<1$, (\ref{dot V}) holds true. Then one can observe that the event-triggered control law (\ref{error bound Zeno free}) would lead (\ref{dot V}) to

\begin{eqnarray}\label{Lyapunov practical stable}
\dot V(x(t)) &\le&  -\sum_{i \in N}(l_i \|x_i(t)\|^2 - j_i \max\{\sigma_i ^2 \|x_i(t)\|^2, \, c_i ^2\} ) \nonumber \\
&\le&  \max \{ -\sum_{i \in N}(l_i-j_i\sigma_i^2) \|x_i(t)\|^2  ,  \nonumber\\
&&- \sum_{i\in N} l_i \|x_i(t)\|^2  +  \sum_{i\in N} j_i c_i ^2 \}  \nonumber\\
&\le&    -\sum_{i \in N}(l_i-j_i\sigma_i^2) \|x_i(t)\|^2  + \sum_{i\in N} j_i c_i ^2 
\end{eqnarray}
which implies practical stability with $\sigma_i < \min\{\sqrt{\frac{l_i}{j_i}}, 1\}$ and finite $c_i$.

Then we introduce the analysis about Zeno-freeness of this distributed event-triggered control law. Since $\dot e_i(t) = -\dot x_i(t)$, then the dynamics of $e_i$ satisfy
\begin{eqnarray}
\dot e_i(t) &=& A_i e_i(t) - \Phi _i x_i(t_k^i) -  \sum_{j \in N_i}(B_iL_{ij}+ H_{ij})x_j(t_k^j) \nonumber\\
&& + \sum_{j \in N_i} H_{ij} e_j(t)
\end{eqnarray}

From the triggering law (\ref{error bound Zeno free}), one can obtain $\|x_i(t_k^i)-x_i(t)\| \le \max\{\sigma_i\|x_i(t)\|, c_i\}$ and further calculations yield $\|x_i(t)\|-\|x_i(t_k^i)\|\le \sigma_i\|x_i(t)\| + c_i $. Thus, it is simple to verify that $\|e_i(t)\| \le \bar \sigma_i \|x_i(t_k^i)\| +  \bar \sigma_i c_i$, where $\bar \sigma_i:=\frac{\sigma_i}{1-\sigma_i}$.

For each $i$, at the instant $t_{k+1}^i$, $\|e_i(t)\|$ satisfies
\begin{eqnarray}
\|e_i(t_{k+1}^i)\| &\le& f_i \|\Phi_i\| \|x_i(t_k^i)\| \nonumber \\
&&+f_i \sum_{j\in N_i}\|B_iL_{ij}+ H_{ij}\| m \nonumber \\
&&+f_i  \sum_{j\in N_i} \|H_{ij}\| \bar \sigma_j (m+c_j)		
\end{eqnarray} 
where $f_i:= \int_{t_{k}^i}^{t_{k+1}^i}e^{A(t^i_{k+1}-\tau)} d\tau$, $m = \max\{\|x_j(t^j_p)\|\}$ for $t_k^i \le t^j_p   <t^i_{k+1}$ and $j\in N_i$. Meanwhile, the triggering law in (\ref{error bound Zeno free}) implies that $\|e_i(t_{k+1}^i)\| \ge c_i$. Then, one immediately sees that
\begin{eqnarray} \label{time bound of zeno}
 \begin{array}{ll}  t_{k+1}^i - t_{k}^i \ge  z_i , \quad\quad\quad\quad\quad \quad \quad \quad \,     \text{if}\,\,\mu_{A_i} \le 0,  \\
		t_{k+1}^i - t_{k}^i \ge  \frac{1}{\mu_{A_i}} \log (z_i\mu_{A_i} +1),  \quad                    \text{if} \,\,\mu_{A_i}>0,  	\end{array} 
\end{eqnarray}
in which
\begin{eqnarray}
	z_i:= \frac{c_i}{ \|\Phi_i\| \|x_i(t_k ^i)\| +  m \sum_{j\in N_i} \zeta_{ij} + \sum_{j \in N_i}     \|H_{ij}\| \bar \sigma _j c_j  }  \nonumber
\end{eqnarray}
where $\zeta_{ij}:= \|B_iL_{ij}+ H_{ij}\|+ \|H_{ij}\| \bar \sigma_j$ and $\mu_{A_i}$ is the logarithmic norm of $A_i$.
Notice that the system is practically stable, so that $\|x_i(t_k^i)\|$ and $m$ are bounded. This implies that $z_i>0$ and hence $t_{k+1}^i - t_{k}^i>0$. 
\qedp

\subsection{Stabilization of distributed systems with hybrid transmission strategy under DoS}
 As a counterpart of Assumption 3, here we assume that there exists a Round-bobin sampling interval $\Delta$ satisfying (\ref{error bound Zeno free}). Now we are ready to present the following result.

\begin{theorem}
	Consider a distributed system as in (\ref{system}) along with a control input as in (\ref{control input}). The plant-controller information exchange takes place over a shared network implementing the event-triggered control law (\ref{error bound Zeno free}) in the absence of DoS. Suppose that there exists a Round-robin sampling interval $\Delta$ such that (\ref{error bound Zeno free}) holds. The network is subject to DoS attacks regulated by Assumption 1 and 2, during which the communication switches to Round-robin until every subsystem updates successfully. Then the distributed system is practically stable if (\ref{Theorem 1}) holds true.
\end{theorem}

\emph{Proof.} 
Similar to the proof of Theorem 1, considering the additional waiting time $N\Delta$ due to Round-robin for the restoring of communications, in DoS-free periods the Lyapunov function satisfies
\begin{eqnarray} 
V(x(t)) &\le& e^{-\omega_1(t-h_n-\tau_n-N\Delta)}V(x(h_n+\tau_n+N\Delta)) \nonumber \\
		&&    + \frac{c}{\omega_1}
\end{eqnarray}
for $t \in [h_n+\tau_n+N\Delta, h_{n+1}[$, where $\omega_1 $ is as in Theorem 1 and $c:= \sum_{i=1}^{N} j_i c_i ^2$. On the other hand, (\ref{Lyaponov in DoS 3}) still holds for $t\in [h_n, h_n+\tau_n+ N\Delta[$. 

Applying the very similar calculation as in Step 3 in the proof of Theorem 1, we obtain
\begin{eqnarray} 
V(x(t)) &\le& e^{-\omega_1   |\bar \Theta (0,t)|} e^{\omega_2   |\bar \Xi (0,t)|} V (x(0)) \nonumber\\
&& + \sum_{ n=0}^{q}  e^{-\omega_1   |\bar \Theta (h_n,t)|} e^{\omega_2   |\bar \Xi (h_n,t)|} \frac{c}{\omega_1} + \frac{c}{\omega_1} \nonumber\\
&\le& e^{\kappa_*(\omega_1+\omega_2)} e^{-\beta_* t} V(x(0)) \nonumber\\
&& +  e^{\kappa_*(\omega_1+\omega_2)}  \sum_{n=0}^{q}  e^{-\beta_* (t-h_n)} \frac{c}{\omega_1} + \frac{c}{\omega_1} \end{eqnarray}
where $n\in \mathbb{N}_0$, $ q:=\sup \{q\in \mathbb{N}_0 | h_q \le t\}$ and $\beta_*$ is as in the proof of Theorem 1. Notice that $t-h_n\ge \tau_Dn(h_n,t)- \tau_D\eta$ by exploiting Assumption 1. Then, the Lyapunov function yields
\begin{eqnarray} \label{Lyapunov function of switching system 2}
V(x(t)) &\le& e^{\kappa_*(\omega_1+\omega_2)} e^{-\beta_* t} V(x(0)) \nonumber\\
&& +  e^{\kappa_*(\omega_1+\omega_2)+\beta_*\tau_D\eta}  \sum_{ n =0 }^{q}  e^{-\beta_* \tau_D n(h_n, t)} \frac{c}{\omega_1}  \nonumber \\
&& + \frac{c}{\omega_1}
\end{eqnarray}
Recalling the definition of Assumption 1, one has that $n(h_n,t)- n(h_{n+1},t)\ge 1$ for $t\ge h_{n+1}$. This implies that
\begin{eqnarray}
\sum_{ n =0}^{q}  e^{-\beta_* \tau_D n(h_n, t)} \le  \frac{1} {1- e^{-\beta_* \tau_D}}
\end{eqnarray}
Finally, (\ref{Lyapunov function of switching system 2}) can be written as 
\begin{eqnarray} 
V(x(t)) &\le& e^{\kappa_*(\omega_1+\omega_2)} e^{-\beta_* t} V(x(0)) \nonumber\\
&& + \frac{e^{\kappa_*(\omega_1+\omega_2)+\beta_*\tau_D\eta} } {1- e^{-\beta_* \tau_D}}\frac{c}{\omega_1} + \frac{c}{\omega_1} 
\end{eqnarray}
If (\ref{Theorem 1}) holds, it is simple to verify that $\beta_*<0$ , which implies practical stability. 
\qedp

\section{Simulation}
\subsection{Example 1}
The numerical example is taken from \cite{forni}. The systems are open-loop unstable such as 
\begin{eqnarray}
\dot x_1(t) &=& x_1(t) + u_1(t) + x_2(t) \nonumber\\
\dot x_2 (t) &=& x_2(t) + u_2(t) \nonumber
\end{eqnarray}
under distributed control inputs such that  
\begin{eqnarray}
u_1(t) &=&  -4.5 x_1(t_k^1) - 1.4 x_2(t_k^2) \nonumber\\
u_2(t) &=&  -6 x_2(t_k^2) - x_1(t_k^1) \nonumber
\end{eqnarray}

Solutions of the Lyapunov equation $\Phi_i ^T P_i + P_i \Phi_i + Q_i =0$ with $Q_i=1$ ($i=1,2$) yields $P_1=0.1429$ and $P_2=0.1$. The matrices are $A=[0.7\,\,\, 0;0 \, \,\,0.9]$, $B=[0\,\,\, 0.0327;  0.1\,\,\,0]$ and $\Gamma=[4.1327\,\,\, 0.4; 0.1\,\,\,   3.6]$ according to Lemma 1. From these parameters, we obtain that the spectral radius $r(A^{-1}B) =0.072 $, $\sigma_1 < 0.3765$ and $\sigma_2 < 0.4657$. We let $\sigma_1 =\sigma_2 = 0.2$. Based on Assumption 3, we choose Round-robin sampling interval $\Delta=0.01$s.

\begin{figure}[bht]
	\begin{center}
		\psfrag{x1}{{\tiny $x_1$}}
		\psfrag{x2}{{\tiny $x_2$}}
		\psfrag{DoS}{{\tiny DoS}}
		\psfrag{ud}{{\scriptsize $u(t_k)$}}
		\includegraphics[width=0.46 \textwidth]{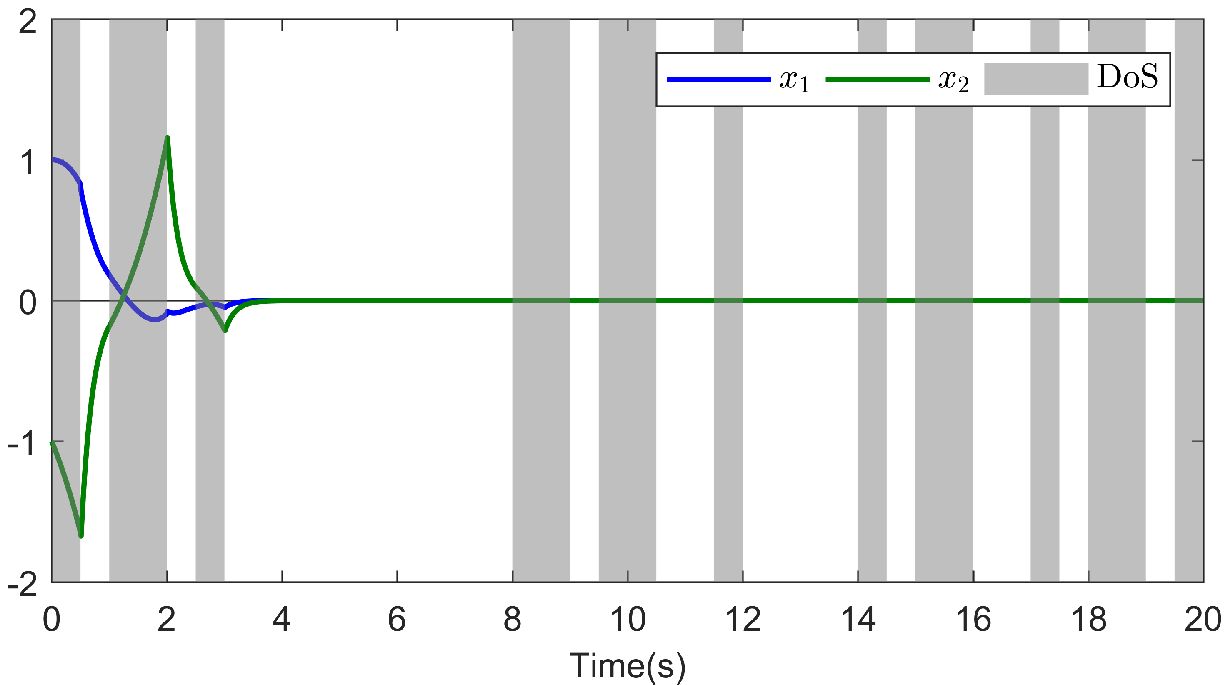} \\
		\includegraphics[width=0.46 \textwidth]{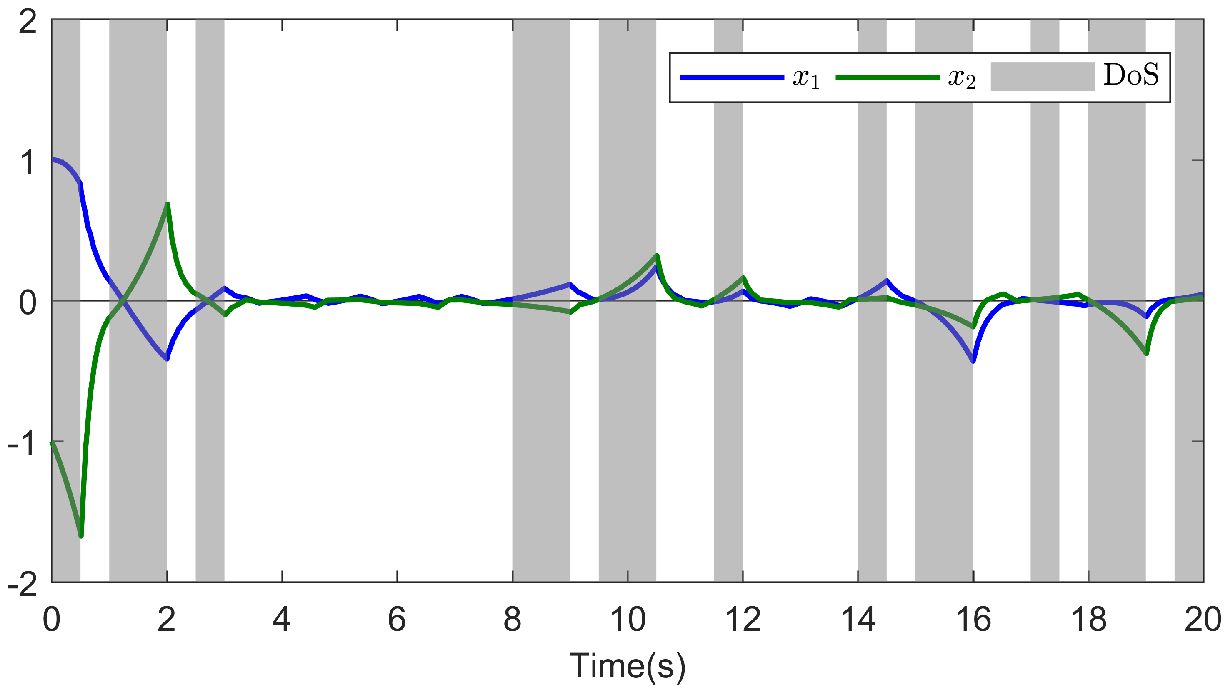} \\
		\linespread{1}\caption{Example 1: Top picture---States under pure Round-robin communication where there are 1200 transmissions in total; Bottom picture---States under hybrid communication strategy where there are 112 transmissions.} {\label{simulation 1}}
	\end{center}
\end{figure}

With those parameters, we obtain the bound $\frac{\omega_1}{\omega_1+\omega_2} \approx 0.0175$ with $\omega_1 \approx 3.0149$ and $\omega_2 \approx 169.3061$.  This implies that a maximum duty cycle of 1.75\% of a sustained DoS would not destabilize our systems in the example. Actually, this bound is conservative. The systems in inspection can endure more DoS without losing stability. As shown in Figure \ref{simulation 1}, lines represent states and gray stripes represent the presence of DoS.
Over a simulation horizon of $20$s, the DoS corresponds to parameters 
of $\tau_D\approx1.8182$ and $T\approx2.5$, and $\sim40\%$ of transmission failures. According to (\ref{Theorem 1}), we obtain 
$
\frac{\Delta_*}{\tau_D} + \frac{1}{T} =0.411 \nonumber
$
, which violates the theoretical bound, but the system is still stable. 

Meanwhile, the hybrid transmission strategy is able to reduce communications effectively. As shown in Figure \ref{simulation 1}. the transmissions with the hybrid transmission strategy is only 10\% of the transmissions with the pure Round-robin strategy.   

\subsection{Example 2}

\begin{figure}[bht]
	\begin{center}
		\psfrag{x1}{{\tiny $x_1$}}
		\psfrag{x2}{{\tiny $x_2$}}
		\psfrag{DoS}{{\tiny DoS}}
		\psfrag{ud}{{\scriptsize $u(t_k)$}}
		\includegraphics[width=0.46 \textwidth]{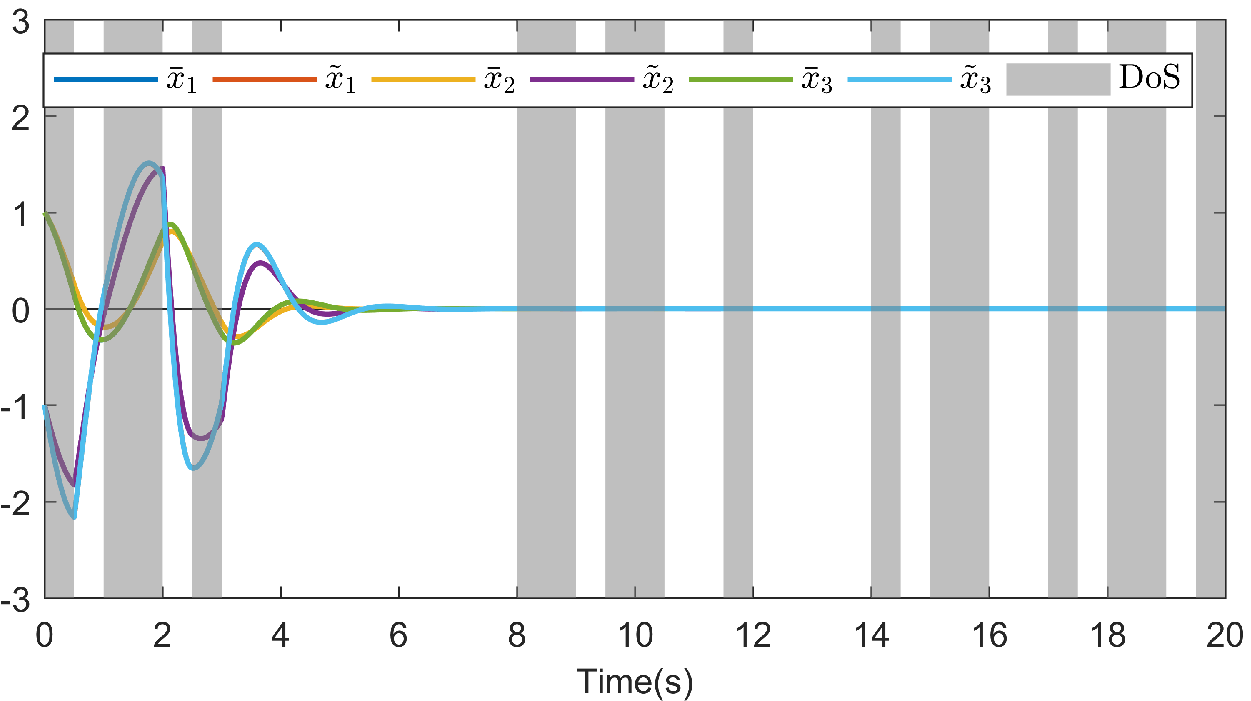} \\
		\includegraphics[width=0.46 \textwidth]{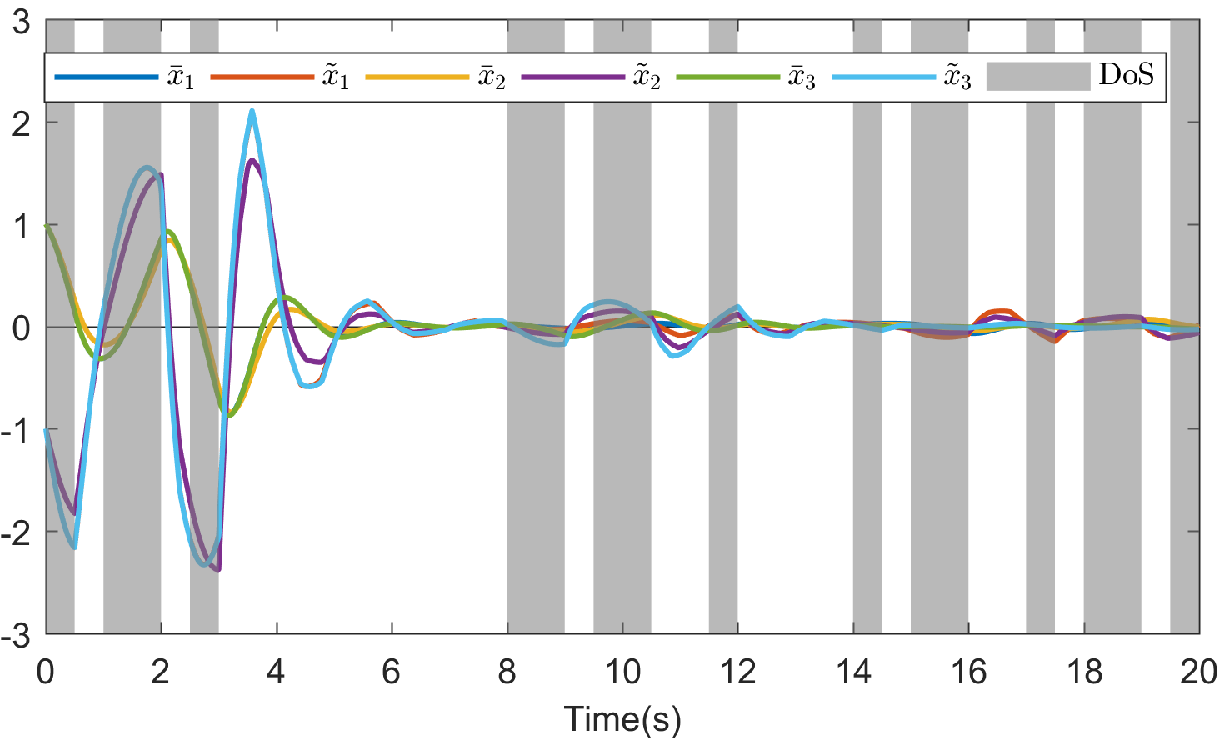} \\
		\linespread{1}\caption{Example 2: Top picture---States under Pure Round-robin communication during which there are 11997 transmissions; Bottom picture---States under hybrid communication strategy where there are 254 transmissions .} {\label{simulation 2}}
	\end{center}
\end{figure}

In this example, we consider a physical system in \cite{wang2008event}. The system is composed of $N$ inverted pendulums interconnected as a line by springs, whose states are $x_i=[\bar x_i, \,\, \tilde x_i]^T$ for $i=1,2, ..., N$. Here, we consider a simple case where $N=3$. The parameters of the pendulums are
\begin{eqnarray}
A_1 =A_3=\left[ \begin{array}{cc}  0 & 1 \\ -3.75 & 0 \end{array} \right], \,\, A_2= \left[ \begin{array}{cc}  0 & 1 \\ -2.5 & 0 \end{array} \right] \nonumber
\end{eqnarray} 
\begin{eqnarray}
B_1 = B_2=B_3=\left[ \begin{array}{cc}  0 \\ 0.25 \end{array} \right] \nonumber
\end{eqnarray} 
\begin{eqnarray}
H_{12} =H_{21}=H_{23}=H_{32}=\left[ \begin{array}{cc}  0 & 0 \\ 1.25 & 0 \end{array} \right] \nonumber
\end{eqnarray} 
The parameter of designed controllers are given by
\begin{eqnarray}
&K_1=K_3=[-23 \,\, -12], \,\, K_2=[-18 \,\, -12]  \nonumber\\
&L_{12}=L_{32}=[-5 \,\,\, 0.25], \,\, L_{21}=L_{23}=[-4.75 \,\, -0.25]  \nonumber
\end{eqnarray}
With the solutions of Lyapunov function $\Phi_i ^T P_i + P_i \Phi_i + Q_i =0$ where $Q_i=I$ and $i=1,2,3$, we obtain
\begin{eqnarray}
A&=&\left[ \begin{array}{ccc}  0.67 & 0 & 0 \\ 0 & 0.45 & 0 \\ 0 & 0 & 0.67\end{array} \right], \nonumber\\ 
B&=&\left[ \begin{array}{ccc}  0 & 0.0608 & 0 \\ 0.1217 & 0 & 0.1217 \\ 0 & 0.0608 & 0 \end{array} \right] \nonumber\\
\Gamma &=&\left[ \begin{array}{ccc}  47.7983 & 24.4007 & 0 \\  22.0276 & 33.2386& 22.0276 \\ 0 & 24.4007 &  47.7983\end{array} \right] \nonumber
\end{eqnarray}
With $A$, $B$ and $\Gamma$ we obtain that $r(A^{-1}B) =0.2216 $, $\sigma_1 < 0.0646$, $\sigma_2 < 0.0844$ and $\sigma_3 < 0.0646$. We select $\sigma_1 = \sigma_2=\sigma_3= 0.01$. The Round-robin sampling interval is chosen as $\Delta=0.001s$ according to Assumption 3. Follow the same procedures as in Example 1, we obtain $\frac{\omega_1}{\omega_1+\omega_2} \approx 0.00012$, which is considerably conservative. In fact, if the systems are under the same DoS attacks as in Example 1, they are still stable, which can be seen from Figure \ref{simulation 2}. The conservativeness is due to the unstable dynamics of the inverted pendulums, the feedback gain $K_i$ and the coupling parameter $L_{ij}$ in the controllers. It is worth investigating how to design suitable $K_i$ and $L_{ij}$ to mitigate this effect (\emph{cf. Remark \ref{remark about final result}}).   

\section{Conclusions}
In this work, we investigated the problem of stabilizing distributed systems under Denial-of-Service, characterizing DoS frequency and duration under which stability can be preserved. In order to save communication resources, we also consider a hybrid communication strategy. It turns out that the hybrid transmission strategy can reduce communication load effectively and prevent Zeno behavior while preserving the same robustness as pure Round-robin protocol. 


An interesting research direction is the stabilization problem of networked distributed systems, where only a fraction of subsystems, possibly time-varying are under DoS. It is also interesting to investigate the problem where DoS attacks imposing on systems are asynchronous with different frequencies and durations. Finally, in the hybrid transmission strategy, the effect of event-triggered control with communication collision can be an interesting direction from a practical viewpoint.



\bibliographystyle{IEEETran}
\bibliography{ref}


\end{document}